\documentclass[10pt,twocolumn]{article}
\pdfoutput=1
\newif\ifOverleaf
\Overleaffalse
\usepackage{times}
\usepackage{fullpage}
\usepackage{algorithm}
\usepackage{algpseudocode}
\usepackage{amsmath}
\usepackage{amsthm}
\usepackage{amsfonts}
\usepackage{amssymb}
\usepackage{array}
\usepackage{appendix}
\usepackage{booktabs}
\usepackage{boxedminipage}
\usepackage{calrsfs}
\usepackage{caption}
\usepackage{color}
\usepackage{enumitem}
\usepackage{fancyhdr}
\usepackage{float}
\usepackage[T1]{fontenc}
\usepackage{graphicx}
\PassOptionsToPackage{hyphens}{url}
\usepackage{hyperref}
\hypersetup{
    colorlinks=false,
    pdfborder={0 0 0},
}
\usepackage[capitalize]{cleveref}
\usepackage[utf8]{inputenc} 
\usepackage{listings}
\usepackage[english]{babel}
\usepackage{lmodern}
\usepackage[activate={true,nocompatibility},final,tracking=true,kerning=true,spacing=true,factor=1100,stretch=10,shrink=10]{microtype} 
\usepackage{multirow}
\usepackage{nicefrac}
\usepackage{pifont}
\usepackage[section]{placeins}
\usepackage{sidecap}
\usepackage{siunitx}
\usepackage{subcaption}
\usepackage{verbatim}
\usepackage[usenames,dvipsnames]{xcolor}
\usepackage{xspace}
\usepackage[ff,sets,adversary,keys,primitives,operators,lambda,logic,notions]{style/cryptocode}

\lstdefinelanguage{javascript}{
  keywords={typeof, new, true, false, catch, function, return, null, catch, switch, var, if, in, while, do, else, case, break},
  keywordstyle=\color{blue}\bfseries,
  ndkeywords={class, export, boolean, throw, implements, import, this},
  ndkeywordstyle=\color{black}\bfseries,
  identifierstyle=\color{black},
  sensitive=false,
  comment=[l]{//},
  morecomment=[s]{/*}{*/},
  commentstyle=\color{purple}\ttfamily,
  stringstyle=\color{red}\ttfamily,
  morestring=[b]',
  morestring=[b]"
}

\newcommand{\squishlist}{\begin{itemize}[itemsep=0pt,parsep=0pt,topsep=0pt,partopsep=0pt,leftmargin=1em,labelwidth=1em,labelsep=0.5em]}
\newcommand{\squishlistend}{\end{itemize}}
\newcommand{\squishend}{\end{itemize}}
\newcommand{\squishenum}{\begin{enumerate}[itemsep=0.5pt,parsep=0pt,topsep=0pt,partopsep=0pt,leftmargin=1.5em,labelwidth=1em,labelsep=0.5em]{}}
\newcommand{\squishenumend}{\end{enumerate}}

\newcommand\myurl[2]{\url{#1}}

\renewcommand{\floatpagefraction}{0.95}

\newcommand{\captionfonts}{\small}
\makeatletter  
\long\def\@makecaption#1#2{%
  \vskip 0.1in
  \sbox\@tempboxa{{\captionfonts #1: #2}}%
  \ifdim \wd\@tempboxa >\hsize
    {\captionfonts #1: #2\par}
  \else
    \hbox to\hsize{\hfil\box\@tempboxa\hfil}%
  \fi
  \vskip 0in}
\makeatother   

\newcommand{\checkpoint}{checkpoint\xspace}
\newcommand{\checkpoints}{checkpoints\xspace}
\newcommand{\commitchain}{rollup\xspace}
\newcommand{\commitchains}{rollups\xspace}
\newcommand{\Commitchain}{Layer 2\xspace}

\newcommand{\executor}{executor\xspace}
\newcommand{\executors}{executors\xspace}

\newcommand{\challenger}{challenger\xspace}
\newcommand{\bridge}{bridge contract\xspace}
\newcommand{\bridges}{bridge contracts\xspace}

\newcommand{\layerone}{underlying blockchain\xspace}

\usepackage{svg}

\newif\ifshowcomment
\showcommentfalse
\newcommand\blfootnote[1]{%
  \begingroup
  \renewcommand\thefootnote{}\footnote{#1}%
  \addtocounter{footnote}{-1}%
  \endgroup
}

\floatstyle{boxed}
\floatname{Workflow}{Listing}  
\newfloat{Workflow}{htb}{lop}

\newcommand{\mytilde}{\raise.17ex\hbox{$\scriptstyle\mathtt{\sim}$}}

\makeatletter
\@ifundefined{showcommenttrue}{\newif\ifshowcomment\showcommentfalse}{}%
\makeatother%
\newcommand{\NewReviewer}[2]{%
\ifshowcomment%
 \expandafter\newcommand\csname #1\endcsname[1]{\textsf{\color{#2}{[#1: {##1}]}}}%
\else
 \expandafter\newcommand\csname #1\endcsname[1]{}%
\fi%
}

\definecolor{bsyColor}{rgb}{0.50, 0.1, 0.8} 
\NewReviewer{bsy}{bsyColor}
\NewReviewer{dawn}{blue}
\NewReviewer{paddy}{cyan}

\ifshowcomment
\newcommand{\todo}[1]{\textsf{\color{red}{[{TODO: #1}]}}}

\else
\newcommand{\todo}[1]{}

\fi

\usepackage[nostamp]{draftwatermark} 

\usepackage{balance}

\ifOverleaf 
\newcommand\githash{OVERLEAF-DRAFT-NO-GIT-HASH\xspace}

\else
\newcommand\githash{09ce\-1681\-ec7d\-3117\-bbe4\-46cd\-6030\-b292\-02e9\-9c3a\xspace}

\fi

\begin{document}

\title{Shades of Finality and \Commitchain Scaling}

\author{
{\rm
  Bennet Yee$^{\dag}$,
  Dawn Song$^{\dag}$,
  Patrick McCorry$^{\ddagger}$,
  Chris Buckland$^{\ddagger}$
} \\
{\small
  $\dag$ Oasis Labs,
  $\ddagger$ Infura
}
} 

\maketitle


\ifshowcomment%
\section*{Reviewer Instructions}
\textbf{%
  {\color{red}DO NOT USE OVERLEAF COMMENTS.} %
  Use \LaTeX-style comments---use the $\backslash$\texttt{NewReviewer}
  macro in \texttt{reviewers.tex}.  Having comments and suggested
  changes in the text makes it possible to synchronize with
  \texttt{github} and edit outside the browser, and having the changes
  in the \texttt{git} history at \texttt{github} rather than squashed
  at overleaf makes it possible to revert previously accepted changes,
  e.g., when the change is no longer appropriate/applicable as a
  result of a later suggestion.}

\textbf{%
  Please pull from \texttt{github.com} to ensure that the overleaf
  copy is fresh (\texttt{Menu > github > pull github changes into
    overleaf}) before making comments.  It would also help if you put
  your comment on its own line, end with a \texttt{\%} to prevent text
  reflows from combining it with the next line to reduce the
  likelihood of merge conflicts, and pushed your changes to
  \texttt{github} after you make them (\texttt{Menu > github > push
    overleaf changes to github}).}
\fi

\section*{Abstract}

Blockchains combine a distributed append-only log with a virtual machine that defines how log entries are interpreted.
By viewing transactions as state transformation functions for the virtual machine, we separate the \emph{naming} of a state from the computation of its value and reaching consensus on that value.
This distinction allows us to separate the notion of \emph{transaction order finality} from \emph{state value finality}.
Further consideration of how blockchain governance handles catastrophic common-mode failures such as zero day exploits lead us to the notion of \emph{checkpoint finality}.

Consensus on the transaction order determines the ground truth.
Everything else---computing the value of a state or handling catastrophic failures such as bugs / zero-day based attacks---are just optimizations.

\paddy{the abstract reminds me of the LazyLedger paper https://arxiv.org/abs/1905.09274 basically a consensus protocol only needs to agree upon the ordering of data and then clients can execute the transactions to determine final state. (they also have a fraud proof / sharded data, not as relevant here).}
\bsy{Thanks.  Added a reference and tried to emphasize that the optimizations, while ``just'' optimizations, are important.}

\section{Introduction}
\label{sec:Introduction}

\blfootnote{The git hash for this paper is \texttt{\githash}.}

The core ideas behind smart contracts and blockchain systems are straightforward, based on a distributed, decentralized append-only log and how we interpret the meaning of those log entries using an abstract virtual machine (VM).

While the basic idea is simple, there are many design choices that need to be made before the full blockchain system is specified, implemented, and becomes usable, e.g., the format of messages that can be logged and their meaning, the VM state machine semantics, etc.
These design choices can have important implications on how the resultant system behaves.

While practitioners understand the nuances, explicating these notions can make it easier for their impact to be discussed and for alternative design choices to be investigated.
This paper attempts to use some basic ideas and notation from programming languages, semantics, and group theory to help model smart contract systems as a way to guide how we look at these design choices, the security issues that arise, and their scaling implications.

We introduce the following distinct flavors of finality:
\begin{itemize}
  \item \textsf{log finality}, when an entry has been irrevocably appended to the blockchain log.
  \item \textsf{transaction order finality}, when a transaction's effect on the VM state is irrevocably determined, without necessarily first computing it.
  \item \textsf{state value finality}, when the computed state value is determined as an efficiently accessed data structure.  This is irrevocable except for recovery from critical infrastructure failures (determined by governance), e.g., an adversary exploits a bug in code used by all participants or an adversary was able to violate a security assumption, such as bribing an above-threshold number of committee participants.
  \item \textsf{\checkpoint finality}, when the computed state value can no longer be changed by hard forks and thus becomes truly irrevocable.
\end{itemize}
We argue that transaction order finality, built using log finality, is the key to reasoning about a blockchain system.
State value finality and \checkpoint finality are nonetheless important optimizations:  they respectively enable application for which independent private state determination is too expensive and permit the storage layer to reclaim storage.

\dawn{ideally we'd like to make the main points and contributions more clearly in introduction.} \bsy{Done.}%

In the next section, we present the ideas, notations, and terminologies that we use to analyze blockchain properties and use them to discuss some well-known blockchain systems and their properties.
Then, in Section~\ref{sec:IdealL2Design}, we discuss what we think are ``ideal'' \commitchain properties and attempt to sketch what such a system would be like.
Finally, we present concluding remarks in Section~\ref{sec:Conclusion}.

\newcommand{\TxTrader}{Trey\xspace}
\newcommand{\TxTraderTp}{he\xspace}
\newcommand{\TxTraderTpp}{his\xspace}
\newcommand{\TxFrontRunner}{Fanny\xspace}
\newcommand{\TxFrontRunnerTp}{she\xspace}
\newcommand{\TxFrontRunnerTpp}{her\xspace}
\newcommand{\eqnGasLimit}{\textsf{gas}_{\textsf{lim}}\xspace}
\newcommand{\eqnGasPrice}{\textsf{gas}_{\textsf{\$}}\xspace}

\section{Key Concepts}
\label{sec:KeyConcepts}

In this section, we introduce ideas, notation, and terminology used to describe and delineate the design space of smart contract systems, using some existing well-known systems as reference points.
Applying these ideas will help us create a simpler, easier-to-understand mathematical / mental model of what all blockchain systems do, and to discuss / analyze trade-offs in current and future designs.
These notions apply both to standard blockchains as well as to scaling solutions such as \commitchains.

\subsection{Append-Only Log}
\label{sec:AppendOnlyLog}

The append-only logs used in Bitcoin and Ethereum v1 are Proof-of-Work (PoW) based designs, whereas those used in Ethereum v2, Cosmos, Polkadot, Oasis,  etc are Proof-of-Stake (PoS) based designs.

\paddy{I feel like there are two parts. e.g. deciding on the data structure for the append-only log. Is it just a chain? a dag? and then the consensus protocol for deciding how to append to the log. the consensus protocol can offer different finality guarantees.}
\bsy{ACID properties will eventually make us provide a serialized view; constructing a DAG, via read/write conflict analysis, yields concurrent execution, but that requires analysis of the transactions (or adding explicit locking) and will require transaction optimization, at a higher level than the log. Not sure we want to get into this here. }
At the level of abstraction needed here, the only thing that we care about is that the append operation for a PoW log is probabilistic, and an entry is not considered successful until about 6 additional entries have been made (the distance to the end of the chain is a security parameter here).
Typically the idea for needing additional entries in PoW is called ``probabilistic finality''.

PoS logs use committee elections and digital signatures instead of solving cryptographic puzzles, and once the consensus protocol completes and a new log entry is made, it is considered final.
While it takes time for the consensus protocol to run, no additional entries are required and this is often referred to as ``instant finality.''

\paddy{There is a concept of "objective" and "subjective" canonical chain highlighted here https://blog.ethereum.org/2014/11/25/proof-stake-learned-love-weak-subjectivity/ the basic premise is that proof of work is "objective" as anyone can download the chain, verify longest chain, and reach the tip. in proof of stake, it is "weak subjective", as the verifier needs to be pointed in the right direction before finding the longest chain. it feels a bit relevant here.}
\bsy{Interesting.  Verifying longest chain requires a sort of global knowledge -- non-existence of other longer chains, which is tricky---it's captured in (ii) for weakly subjective definition, but any time there is a $\forall$ is complicated. }
We will refer to the general idea as ``log finality,'' whether probabilistic (with appropriate security parameter) or instant.
Log finality is the mechanism upon which other notions of finality is built: log finality only says that some entry is successfully appended to the log, but says nothing about what that entry means.

\dawn{usually we don't consider PoW has log finality; and we can say that in practice this is what people use. also, instead of log finanilty, should it be called transaction finality? or transactin log finality?} \bsy{Log finality is not transactional since no ACID transaction execution / interpretation of the messages has occurred.}

\subsection{Virtual Machine State}

Since ``state'' can be encoded and represented in many ways, we need to start with mathematical necessities to describe what we mean by ``state'' in a way that is divorced from any particular representation.
By ``state'', we mean a function from a key set, its domain, to a value set, its range, i.e., $\textsf{State}: \textsf{Key}\rightarrow\textsf{Value}$.
In Bitcoin, \textsf{Key} would essentially be public keys, and \textsf{Value} would be numbers representing the quantity of Bitcoin tokens under the control of the corresponding private keys.
In Ethereum-like systems, \textsf{Key} would be a union of several disjoint sets, e.g., public key hashes associated with Externally Owned Accounts (EOAs), contract addresses, contract address and 256-bit address tuples for naming persistent store locations, etc.
Correspondingly, \textsf{Value} would be public keys, EOA balances, contracts' code and balance, and 256-bit values from contracts' persistent store, etc.\footnote{This is just one way to model Ethereum state; other models, e.g., as a set of mappings one for each key type, would be more precise.}

\subsection{Transactions As Curried Functions}

Users of blockchain-based systems propose transactions that are recorded by being appended to the blockchain.
We normally think of transactions as altering state.
In a mathematical description, they are functions that \emph{transform} state, i.e., $\textsf{State}\rightarrow\textsf{State}$.
Smart contract entry points take arguments---in Ethereum, the callData---and does not fit this type signature; however, by currying all the user-supplied arguments including authorization information (\texttt{msg.sender}, etc), all transactions can be modeled in this manner.\footnote{Not all state transformations are valid transactions, e.g., a state transformation that changes a contract's state in violation of its code invariants.}

Depending on the smart contract system and the type of analysis we need to do, we may need a model that exposes more structure of transactions.
For example, it is useful to model an Ethereum(-like) transaction $t$ as two component functions: first, $t_c: \textsf{State}\rightarrow\textsf{State}_{\bot}$, representing the call into a contract entry point, returning an updated state modeling the effects of the contract code execution on contract persistent storage (no change if the transaction aborts, with $\bot$ representing non-termination); and second, $t_g: \textsf{State}\rightarrow\mathbb{N}_{\bot}$ representing the gas consumed from execution (until either commit or abort), returning $\bot$ if gas limit would be exceeded (including non-termination).
The function $t$ is then the following composition of $t_c$ and $t_g$, with $a_b$ representing the key for the balance for account $a$ (which is \texttt{msg.sender} for $t$) and store update notation from semantics:
\[
\small
    t(s) = \begin{cases}
        [a_b\mapsto s(a_b)-\eqnGasLimit\cdot\eqnGasPrice] s & \text{if $t_g(\sigma) = \bot$}, \\
        [a_b\mapsto t_c(s)(a_b)-t_g(s)\cdot\eqnGasPrice] t_c(s) & \text{otherwise}
        \end{cases}
\]
This is a useful separation since most contract virtual machines implement standard ACID transaction semantics, and an explicit transaction abort reverting the contract state (failure atomicity) is just $t_c$ returning its input state, though gas fees for computation performed until the abort must still be paid.

\subsection{Natural Names of State}

Because users think of their transactions as being executed in a strict serial order and submit transactions based on their model of what the current system state (e.g., their EOA balance), a natural way to refer to any reachable state is the sequence of transactions that yields that state.

\subsubsection{Names And Composition of Transformations}

Transactions are state transformations, and transforms form a group under function composition.
We use the ``;'' function composition notation from programming language semantics, so we write the state after the $j^{\text{th}}$ transaction as $s_j= (t_1;t_2;\cdots;t_j)(s_0)$.
For a state $s$ that is reachable via a sequence transaction from genesis, we call ``$t_1;t_2;\cdots;t_j$'' (one of) its ``natural name(s).''

Note that for Bitcoin, Ethereum, and Arbitrum, the natural name is exactly what is recorded on the blockchain.
Other smart contract systems may only \emph{implicitly} record the execution order, e.g., as part of a zkSNARK proving correct execution.

\subsubsection{Names Are Not Unique}

Note that names are not unique.
Many states have multiple names.
If two transactions $f$ and $g$ commute at a given state $s$, i.e., the resultant state is $(f;g)(s) = (g;f)(s)$, then there are (at least) two names for the resultant state.
Of course, if both $f$ and $g$ were proposed, the system will eventually decide on \emph{some} order.
This is analogous to how $0+1+2$ and $0+2+1$ are both ways to name the value $3$.

Note also that some states have no names in this nomenclature.
For example, a state where an EOA controls more tokens than the total token supply is unreachable, since (absent a catastrophic bug) there are no sequences of transactions from the genesis state that would result in such a state.
This is fine, since such unreachable states are of no interest.

\subsection{Ground Truth}

Ground truth for smart contract / blockchain systems is knowing what the current state is supposed to be.
This does not mean, however, that we must have an efficient representation of that state, merely that it is computable---efficiently, and unambiguously.
We argue that the most natural definition for blockchain system state is based on consensus agreement on the order of all transactions since the genesis state (or since a \checkpoint state, see Section~\ref{sec:Checkpoints}).

Once a transaction's parameters are logged (has achieved log finality), its execution order relative to earlier logged transactions is determined.
Via a simple inductive argument, since the input state is computable by the execution order finality of earlier transactions, the resultant state from this transaction can, in principle, also be computed.
We call this ``transaction order finality.''

Consensus about the current state is based on states' natural names, and computing an efficient representation is not necessarily coupled with this decision.
Everything else is an optimization.

\paddy{Going back to the Mustafa paper. An 'execution' of a transaction is only important client-side. In the case of commitchains/rollups, the bridge contract is the client for the other system. That is why the 'execution result' is important as it results in some side-effect. i.e., issuing a withdrawal.}
\bsy{Yes, need to talk about side effects / externally visibility. Added.}

\subsection{Efficient Representations of State}
\label{sec:EfficientRepresentationsOfState}

All blockchain-based systems use an append-only log to record ground truth.
From those log entries the system determines the consensus reality on the state of one (or more) virtual machines.
However, different blockchains handle state differently.

Bitcoin only records transactions in the blockchain blocks.
The VM is relatively simple and the state is the number of tokens controlled by public keys---the so-called Unspent Transaction Outputs (UTXOs).
The logged messages specify how to transfer tokens using the VM rules, but state is implicit.
In contrast, Ethereum also records a representation of the new state that results after all the transactions named within the block are applied to the state from the previous block.
This state is recorded via a cryptographic commitment---the ``stateRoot'', which is the root hash summary of the Merkle Patricia trie representation of the Ethereum Virtual Machine (EVM) state.
Hyperledger~\cite{albassam2019lazyledger} takes a Bitcoin-like approach and leaves the interpretation of results---the effect on the VM---to clients. 

Note that Bitcoin miners must also have an efficient representation of state, since in order to verify new blocks double spending checks etc must be performed---it is just that this state does not show up on-chain.
Availability is separate from permanence: in either case, miners has to either ``catch up'' by replaying transactions since a \checkpoint (more on that later) or obtain a copy from somebody else.
Apart from replaying transactions, there is no way for Bitcoin miners to verify state obtained from an untrusted party; Ethereum miners, on the other hand, can readily verify the Merklized data structure (stateRoot).

Rollup systems use an existing \layerone as a security anchor to allow ``off-chain'' execution, where smart contract code executes in a separate VM (``\commitchain''), so that many transactions can execute there while requiring fewer (or cheaper) transactions at the \layerone.
This is an attractive scaling solution, since (presumably) computation on the \commitchain is cheaper, and a single \layerone can support multiple \commitchains.
Optimistic rollup designs like Arbitrum~\cite{kalodner2018arbitrum} commit the transaction order to a \bridge in the \layerone, computes the resultant state in the \commitchain, and sends that result to the \bridge in the \layerone to validate.
We will discuss various validation schemes later (see Section~\ref{sec:StateDetermination}); the key observation here is that transaction order can be determined in a separate step from state computation.

Another way to think of this is that the resultant state from the execution of a transaction $t$ is \emph{named} once we have the order of all transactions starting from the genesis state up to and including $t$.
We may not yet have \emph{computed} that state as a data structure that allows the key-value lookup to be performed efficiently, but there is no doubt what that state would be---everyone given its name will arrive at the same abstract mapping, even if its concrete representation---actual choices of data structures---may differ.
Bitcoin only names the current state.\footnote{Though states are realized periodically, when \emph{\checkpoints} are created.  See Section~\ref{sec:Checkpoints} for more details.}
Ethereum-like systems both name and compute one particular representation of state.
Some rollups separate naming from state computation, where the transaction order is committed to the layer 1 blockchain first, and the correct resultant state is computed off-chain, in the layer 2 \commitchain, and committed to layer 1 at a later time.
Other rollups use a mempool design for the \commitchain, so that the transaction order is determined by \commitchain{} nodes rather than in the \layerone, and that order is committed along with the computed state hash.
Such a design reduces the number of transactions in the \layerone, trading off reduced transaction processing cost there for tight coupling between transaction order finality and state determination.

The separation of concerns should be explicit.
The consensus layer is responsible for making an immutable, append-only log.
Its primary purpose is to record the transaction---the calldata---in the order in which they are to be executed.
\emph{This names the resultant state, and everything else is secondary.}
Computing and agreeing on this state can come later, as fulfilling a promised value.
The result---a correspondence between a named state and the state representation---can be similarly recorded to make state representation sharing easier.
That is to say: Bitcoin logs only execution order; Ethereum logs both execution order and an efficient representation of the resultant state in a tightly coupled manner; decoupled layer 2 rollup designs, on the other hand, logs both but separately, so that problems with the validity of the computed state does not necessarily invalidate the committed transaction order.

Note that it is important to calculate state and to reach consensus on it, since blockchains are not closed systems:  transfers between \commitchain and \layerone is an example of external actions that depend on state; in general, any contracts that can cause off-chain effects, such as the shipment of goods, require consensus and state value finality.

In our view, obtaining an equivalent but more efficient representation of a given state is an important optimization.
Indeed, committing the transaction execution order yields one sense of finality---transaction order finality, allowing us to name a committed state.
Computing the stateRoot and reaching agreement on the result yields a second sense of finality---``state value finality'', allowing us to confidently \emph{use} the state's value or representation.

\subsection{Checkpoints}
\label{sec:Checkpoints}

Replaying all transactions from genesis is expensive and as a blockchain system is used it quickly becomes prohibitively so for new participants wishing to join the system.
The storage cost of the blockchain blocks grows linearly over time, and for Ethereum maintaining any sort of availability for old trie state can become quite expensive.
In the past, Bitcoin performed a sort of uber-commit by releasing new clients with ``checkpoints'', where the state at some block height is built into the client.
Blocks with lower block numbers can be safely discarded, since everyone has a state---the checkpoint state---from which to replay newer transactions.
Bitcoin has since removed checkpoints because it was viewed as creating confusion/misconceptions around the security model.

\paddy{It might be worth double checking Bitcoin checkpoints. https://bitcoin.stackexchange.com/questions/75733/why-does-bitcoin-no-longer-have-checkpoints as far as i'm aware they were removed due to misconceptions around the security model.} \bsy{Thanks---added discussion of insurance.}
In addition to reducing the entry cost for joining the system, checkpointing reduces the cost of record keeping for existing participants.
This is a ``meta'' level of finality: the effect of transactions that landed prior to the \checkpoint cannot be disputed, since the records associated with them are likely to be unavailable.

Beyond cost of entry or on-going operational costs, \checkpoints are also used in conjunction with governance mechanisms for catastrophic error recovery.
Many blockchain systems that have experienced problems, e.g., massive token losses due to bugs in critical blockchain code or smart contracts, have resorted to hard forks to recover from such errors despite transactions reaching state finality; checkpointing too soon---if records older than \checkpoints are not kept around---would prevent recovery by reverting to the \checkpoint state and (optionally) re-executing (using fixed versions of the code, etc) transactions in their original logged order.

\subsubsection{Checkpoints vs Long-Range Attacks}

Note that \checkpoints addresses a different problem than long-range attacks.
In long-range attacks, we are worried about exposure of old cryptographic signing keys used by past consensus committee members in a PoS design~\cite{azouvi2020winkle}.
Such members may have exited the ecosystem and their old keys are no longer handled carefully; worse, such past members may rationally decided to auction their signing keys on the dark web, since they no longer hold any tokens and have nothing to lose.
Such keys are useless when used to present bogus information to blockchain participants that know the current committee composition and have been tracking transactions and committee elections.
However, consider a threat model where a Rip van Winkle victim wakes up after a period of inactivity and is somehow placed in an \emph{Inception} style virtual world / faux information bubble.
That bubble filters out legitimate information about a PoS blockchain's current state and instead only makes available information constructed to make a forked chain---made feasible due to a super-threshold number of exfiltrated/compromised keys of members from an old consensus committee \cite{azouvi2020winkle}.

The Inception information bubble is an interesting threat model.
If applied to PoW blockchains, a victim cannot know if the chain that they see is indeed the longest chain---the ``longest'' predicate amounts to a universal quantifier, and global information is needed.
Estimates for how long the chain \emph{should} be might be feasible if there is trusted time, but that's probabilistic in nature: block production rate depends on both protocol parameter changes and the number of active miners, which has more to do with the economic attractiveness for participating (relative to all other investments) than with computation power limitations.
Furthermore, such a length estimate only applies to the whole chain segment since the victim fell asleep and does not help much with the ``longest'' predicate since the fork can be quite recent.

Without Inception-like powers to mount eclipse attacks~\cite{190890}, an adversary should be unable to confuse potential victims.
A potential victim can verify that they have a current view of the blockchain: they just securely query $N$ sources for the hashes of recent blocks on the chain.
If a majority $M$ of these hashes are on the same chain and these sources are honest, not eclipsed, and have continued to be blockchain observers , the potential victim will be able to distinguish the global consensus chain from a forked chain created with long-range leaked keys.
Here $M \le N$ is a security parameter, which can be chosen so that the Inception-esque adversary will have to additionally compromise prohibitively many more keys (or their holders) than just those of old consensus committee members, since any blockchain observer can witness recent blocks and thus $N$ can be much larger than the size of a consensus committee.

\subsubsection{Zero-day Attacks / Common-mode Failures}

Checkpointing is intended for addressing the handling of relatively \emph{recent} zero-day attacks where a super-threshold number of the current consensus committee members have been compromised, or a newly discovered vulnerability in the \commitchain software is being exploited.
We do not envision it being useful for handling attacks that had not been noticed for a long time, since any actual means of addressing it will be complicated.
The cascading causal relationship of newer transactions depending on the output states of older transactions is likely to lead to an explosion of transactions that will be aborted in a new interpretation of their effects when they had successfuly committed before.

The design decision is whether to perform checkpointing at all, and if so, how old---in real time, block numbers, etc---must a finalized transaction be before it might be included in a \checkpoint.
This decision is essentially the blockchain version of a \emph{statute of limitations} in many legal systems.
Unlike normal statutes of limitations that specify a time limit that is specific to the type of crime---and for some crimes there are no limits---here the \checkpoint is global in scope:  all transactions, regardless of which contracts they might be associated with, have to be treated the same way.

In our mathematical model, deciding / making a \checkpoint is essentially choosing a state and ``hardcoding'' the association between its natural name (or block number) and a specific value in a common, efficient representation.
The treatment of \checkpoints is analogous to how the genesis state $s_0$'s value is hard coded in the system.
Periodically, a state $s_{c_j}$ that has been long finalized is chosen to become a \checkpoint, where $c_0=0$, $c_{j} \lll c_{j+1}$, and $s_{c_{j+1}}=(t_{c_j+1};\cdots;t_{c_{j+1}})(s_{c_j})$.
For all intents and purposes, the \checkpoint state behaves like the genesis state; after the \checkpoint creation, state names can be specified relative to the \checkpoint.

The \checkpoint's state / value association is a temporal barrier: transactions earlier than the \checkpoint have ``\checkpoint finality'', since records that would enable their replay in the alternate bug-fixed environment are unavailable.
Note that unlike transaction order finality and state finality which typically occur at the same time as when their log entries are made, i.e., at log finality, \checkpoint finality could occur long after the identification of a state as a \checkpoint candidate.
For example, a system could log that a state will become the next \checkpoint once the blockchain's block height reaches a certain value (which is expected to occur in about six months, say) or when a quorum of time oracles attest that a certain date has been reached.
The log entry makes the decision irrevocable, but \checkpoint finality does not necessarily occur until other gating conditions are met.

Separating state finality from \checkpoint finality to handle the possibility of catastrophic failures introduces risk for those who need to take actions external to the \commitchain based on state finality.
External actions cannot always be rolled back.
We believe that this can be handled using an insurance model: for example, an insurance policy based on the type of external action and risk profile could be offered to make participants whole, should a checkpoint/replay invoked due to a catastrophic failure cause the proper external action to change.

\subsection{Execution Parameters: callData}
\label{sec:ExecutionData}

When we name states via transaction order, it is a ``natural'' name as a sequence of state-transformation functions as applied to the genesis or a \checkpoint state.
The state-transformations must be fully specified, i.e., the transaction data (callData) are function parameters to the transaction call that, when curried, allow us to view the transaction as state transforms.

In order for the commitment of transactions to make sense, it must include all callData for the state to be computed.
This is a \emph{data availability} requirement.
Transactions (their callData) does not have to be directly in the blocks, though that is the simplest design choice.
If a highly available data repository, i.e., \emph{a data availability layer} (e.g., IPFS~\cite{benet2014ipfs}), can be used, then lists of transactions can be stored there and the on-chain data can simply be the name by which the transaction list can be obtained.

Here, feasibility must include some notion of verifying data availability, e.g., signatures from a quorum of availability providers.
It is not enough to use a simple content-addressable storage (CAS) without availability guarantees, since otherwise we face the following dilemma when an adversary computes the CAS name without actually making the data available.
In such a scenario, we either sacrifice liveness, having to wait indefinitely for the callData to become available, or try to use timeouts and treat unavailable callData as implicit aborted transactions and move on.
This latter choice sacrifices finality: a user can nullify their transactions after the transaction's execution order has been decided---by making their callData unavailable---if the user decides that the execution order is not to their advantage, so until state finality has been reached, transaction order finality does not uniquely specify a resultant state.

Note that Ethereum's stateRoot has a symmetric CAS availability problem.
The state output from a transaction is needed as input to the next transaction, and a lack of availability here means that the new state's representation cannot be easily computed, destroying liveness.\footnote{Obviously the input state can be reconstructed by transaction replay, but that is not efficient.}

\subsection{Execution Ordering}
\label{sec:ExecutionOrdering}

Execution ordering can be quite important, since any given two state transforming functions might not commute with each other.
The notion of ``front running'' is not new with blockchains but has been unethically (and illegally) practiced in stock and commodities markets~\cite{InvestopediaFrontRunning}. 
Front-running uses the ability to influence execution ordering results, where additional orders are injected in front of (and often also behind) orders that may move the market.
For example, if \TxTrader submits a transaction $g$ to buy a large amount of some commodity, it is likely to cause a price increase (``move the market'').
If \TxFrontRunner knew about \TxTraderTpp order and can inject in a pair of transactions to sandwich \TxTraderTpp: the net is a \emph{pseudo-conjugate transaction} $g' = f;g;f^{*}$, where $f$ denotes a transaction to buy the same commodity at market price, and $f^{*}$ denotes a ``pseudo-inverse'' transaction to sell the same amount, at what will be a higher market price, to exit the market and reap profits.\footnote{
  The notion of conjugacy from group theory captures the idea of frontrunning, $f^{*}$ is not the inverse because transactions are not invertible: gas fees must be paid.
  Furthermore, the transactions occur at the market price and the portion of state representing the token balance for \TxFrontRunner would not return to its original value---the point of front-running is to make money!
  Because of this, we say that $f^{*}$ is a ``pseudo-inverse.''
  Additionally, if \TxFrontRunner wants to hide \TxFrontRunnerTpp activities the trailing transaction could, for example, sell a smaller amount to reap most of \TxFrontRunnerTpp profits without completely exiting the market, making identifying and matching the leading and trailing transactions difficult.
}
In Bitcoin, Ethereum, and most layer 2 rollup designs, the (layer 1) miners choose transactions from a mempool of proposed transactions and can choose and order transactions within the new block in any order that they want.

Ideally, transactions that offer approximately the same gas fees should perhaps be handled on a first-come, first-serve basis, with the order decided based on approximate timestamps.
Standard Bitcoin, Ethereum, and rollup designs are all vulnerable to order conjugation, though there are research on ways to maintain fair ordering for some definition of fairness~\cite{kelkar2020order,MEV,9152675}.

How to decide on an execution ordering is a largely orthogonal design decision.
The scheme used in Bitcoin and Ethereum (v1) is leaderless and probabilistic, since miners can independently choose the transactions to include in the blocks they mine.
For L2 rollups with decoupled ordering and state value determination, the transaction order can be determined in many ways: (1) as a ``null hypothesis'', in L1 submission order (susceptible to L1 front running); (2) via trusted off-chain schedulers that (fairly) order batches of transactions; (3) via trustless decentralized applications on other blockchains that perform batch scheduling, using commit-reveal of transaction details to prevent biased ordering; etc.
The potential design space is large, and exploring this is on-going research.

\subsection{Deciding on the Correct State}
\label{sec:StateDetermination}

The mechanism by which the correct state is decided is security critical.
There are many different design choices that have been explored, with different security assumptions and scaling efficiency.
By security assumptions, we mean not only common cryptographic assumptions (e.g., one-way functions), but also compute resources such as hashing power and monetary resources such as control of some fraction of the total stake.
By scaling efficiency, we mean both in terms of normal resource usage and in terms of resources needed to achieve a desired level of security.
Table~\ref{tab:L2SystemsTable} shows several different rollups and their state transition security mechanisms.

\begin{table}[tpbh]
\footnotesize
\begin{tabular}{|l|p{1.35cm}|p{1.75cm}|p{1.5cm}|}
  \hline
  \textbf{Rollup} & \textbf{Base chain} & \textbf{Security} & \textbf{VM Types} \\
  \hline
  \hline
  Arbitrum & Ethereum & Multi-round fraud proof & EVM/AVM \\
  \hline
  Oasis & Oasis consensus layer (Tendermint) & Bare-metal fraud proof & Rust (Wasm), EVM, (extensible)\\
  \hline
  Optimism & Ethereum & One-round fraud proof & EVM \\  
  \hline
  StarkEx & Ethereum & Validity proof & Swaps \\  
  \hline
  ZKSync & Ethereum & Validity proof & Transfers \\
  \hline
\end{tabular}
\caption{Comparison of Different Rollup Designs}
\label{tab:L2SystemsTable}
\end{table}

\subsubsection{Proof of Work}

While Bitcoin uses a PoW-based log, there is no explicit decision to be made about the correct state since no state is recorded.

\paddy{I've been thinking about the Bitcoin state. In UTXO, the state is the list of unspent outputs and each transaction references the hash of the previous output. So because hashes are used to reference an output, that should imply some strict state implementation? i've not seen too much mentioned about it.}
\bsy{Yes.  I haven't seen an explicit state machine description for UTXO---it's always been kind of implied AFAIK.  The hashes are just a kind of back-pointers to state, which is an amount of tokens transferred and the Bitcoin script that gates whether they can be spent by future transactions.}
In Ethereum v1, PoW is used and hashing is the resource bottleneck.
All miners are validators, and the security assumption is that malicious actors cannot control more than half of the hashing power or mount a ``51\% attack.''
Here, validation means re-executing the transactions to verify that the new stateRoot matches.
Invalid blocks are ignored and not considered as extending the chain.

One issue is that even though it appears that all miners independently validate transactions, the incentives for rational miners are to behave otherwise.
Because the execution order and stateRoot must be placed in the same block and thus are tightly coupled, there is a performance advantage for small miners to join together and form a mining pool/cooperative.
Rather than individual miners re-executing transactions from the block to compute the stateRoot, a mining pool could compute the stateRoot once on the pool's fastest machine, and then devotes all of their compute power to solving the hash puzzle.
This obviously also centralizes what would have been replicated transaction executions, robbing the system of the degree of independent verification represented by the mining pool members.

This is not (yet) an acute problem, since hashing costs dominate that of VM execution to process the transactions, and the savings from sharing stateRoot is not significant.  The incentive to do so increases if/when the gas limit becomes high enough so that VM execution cost becomes noticeable.

\subsubsection{Proof of Stake}

Ethereum v2, Cosmos, etc are PoS based blockchains.

All consensus committee members are also validators, and the security assumption is that malicious actors cannot amass more than a supermajority (usually 2/3) of the total stake.
Validators earn block rewards.
Stake-based voting and reward distribution remove most of the incentives to mount Sybil attacks, but does not address other attacks such as exploiting zero days, which are common-mode faults.

\subsubsection{ZK Rollups}

Zero-knowledge (ZK) rollups use zkSNARKs to prove the correctness of the \commitchain{} virtual machine execution.
Because security proof verification is involved, a single honest verifier is enough to keep the \executor/prover honest.

The key idea is that proof verification is very cheap; thus, the security parameter for the number of validator---proof verifiers---can be easily increased.
Unfortunately, proof generation is relatively expensive, so while payment transactions are feasible (ZCash~\cite{hopwood2016zcash}), the case for general smart contract computation is more difficult.
Proof generation for Ethereum-style computation is being developed/researched (zkEVM~\cite{zkEVM}).
It remains to be seen how good a scaling solution this represents, since the \commitchain where the proof generation occurs may be slow/resource intensive, so even though the \layerone might be able to run \bridges for many ZK rollup \commitchains, the aggregate throughput and efficiency may still be insufficient for general, complex applications.

The proof verification algorithm takes time to run.
This should be done within the \bridge, and the typical description of ZK rollups is that an invalid proof is treated as if no proof were submitted, i.e., \bridge aborts the faux proof submission transaction.
As long as there is enough replication in the \layerone so that the consensus there reflects the correct execution of proof verification, the security of the \commitchain is guaranteed.

\subsubsection{Optimistic Rollups}

In optimistic rollups, validators are any entity that can re-execute the transactions and compare the resultant state.
An \executor commits its result as a Disputable Assertion (DA) to the \layerone, and any validator that finds a discrepancy can issue a challenge (and the validator becomes a \challenger).
After dispute resolution, if the DA is found to be incorrect, the \executor is slashed and the successful \challenger earns a reward.
This means a single honest validator is enough to keep the \executor honest.

Validators must be given time to re-execute the transactions and to generate challenges.
Because re-execution is more expensive, this can create a significant delay for state finality: a DA is only considered accepted when the disputation period has passed.
The appropriate length of the disputation period depends on the maximum gas allowed in a rolled up batch of transactions and other factors; current designs allow as much as a week.

In Arbitrum, users are expected to not have to wait for the disputation period, based on the notion of ``trustless finality''.
The scenario is that a user has submitted a transaction and it is in-flight: a DA includes it, but the disputation period has not yet passed.
The user wishes to propose a new transaction that depends on its result.
If there are doubts about the DA, the user might be hesitant to propose the new transaction.
The argument is that since the user could act as a validator and re-execute all the transactions leading up to the DA and thus independently validate their mental model of system state, they should be able to freely submit their new transaction proposal.
This is essentially using transaction order finality: users trust the \layerone's recording of the \commitchain transaction order and that the VM is defined and implemented correctly to be deterministic.

Another argument is because having multiple \executors/validators is feasible.
Here, they can post additional DAs (for new transactions) that depend on earlier DAs that have not yet reached state finality.
These additional DAs provide additional assurance that the earlier DAs are correct, since presumably the new dependent DAs have validated the earlier DAs.
Currently, Arbitrum has a single centralized \executor, though anyone can be a validator/\challenger.

Verifiers suffer from the verifier's dilemma~\cite{luu2015demystifying,verifiersdilemma}.
Even though proof verification might be cheap, it is cheaper still to assume that somebody else has done the verification, when the odds that an \executor will try to cheat appears to be low.

\subsubsection{Bare-Metal Fraud Proofs}

The bare-metal fraud proof approach is a traditional fast-path / slow-path approach seen in systems design, where the fast path efficiently handles the common case, with a slower path that acts as a fallback for the uncommon case(s).
Here, a single honest node in the fast path suffices to trigger slow-path verification.
It as agnostic with respect to the \commitchain VM, since there is no need to single-step VM execution to find exactly where an error occurred.
To understand the bare-metal fraud proof approach, it is useful to understand the system architecture.

The bare-metal fraud proof approach can be viewed as a form of optimistic rollup, but with a committee of nodes executing in parallel verifying each other rather than taking the executor vs challenger view.
The VM execution of smart contracts is separated from consensus.
There are two types of committees: a consensus committee executing a consensus protocol, and one or more compute committees executing the VM---similar to rollup executors running smart contracts off-chain---doing the heavier lifting of general smart contract execution.
The consensus layer records the resultant VM state from the compute committee.

Unlike the consensus committee which runs a single consensus protocol, we view the compute layer as running two protocols:  the fast-path protocol of \emph{discrepancy detection} (DD), where we can use a smaller sized primary committee; and the slow-path protocol of \emph{discrepancy resolution} (DR), where a (much) larger backup committee would be used.
The goal, which is justified by the security parameter calculations (see Appendix~\ref{sec:DDParameters}), is to use small primary committees without compromising security.
This is the source of the efficiency improvement of the bare-metal fraud proof approach: the amount of replicated computation is significantly reduced in the fast path, with an incentive design such that the slow path should never be used.

Note that transaction ordering is an orthogonal design decision from the use of DD/DR.
It can be decoupled and committed to the consensus layer as separate log entries prior to smart contract execution, or a leader from the compute committee could be selected (possibly rotating) to choose an execution order from the available transaction proposals.

However execution schedules are determined, all members of the compute committee run in parallel and determines the state that would result from the execution of a batch of transactions.
Compute committee members sign their computed output states.
Equivocation is punished by slashing stake like in many consensus designs, and we will assume that each committee member will sign at most one state as the result from the execution of a batch of transactions henceforth.

The key idea behind DD is that when all compute committee members agrees on the resultant state, we can commit it to the log, and 
this is secure as long as the size of the primary compute committee has enough members so that the probability that \emph{none} of the reporting members are honest is negligibly small.
If there is any disagreement among the primary committee member, we do not know which of the reported output states is correct---this is the key difference from other state validation schemes:  DD only performs error detection and not error correction.
Instead, we switch to the DR protocol with the larger backup committee, and \emph{that} output state is deemed to be the correct one.
In essence, the DR protocol is a parameter to the bare-metal fraud proof approach: the DR protocol can use a much larger committee---even all available nodes---and use much more resources.
Because those DD nodes that reported a differing output state from that determined by DR will have their stake slashed, there is no incentive to deviate from the protocol unless the adversary can either control the entire DD committee or enough of the backup committee for DR to fail.

Note that the choice of DR does \emph{not} have to only involve larger committees: we could use a bisection algorithm to find the single VM instruction at which the computation of the two (or more) resultant states first diverged~\cite{teutsch2019scalable,kalodner2018arbitrum}.
While this approach is great from an algorithmic standpoint, it is challenging to implement since the entire VM instruction set must be re-implemented in the \bridge to see which instruction executed incorrectly and thus resolve the discrepancy.
Maintaining VM agnosticism allows the system greater flexibility: a VM can be designed to allow the VM programs to be compiled to (sandboxed) native code, allowing applications such as data analysis that is not currently feasible on blockchains or \commitchains.

The consensus layer accepting and committing a state from DD/DR yields state finality.
Just as an exploit using zero-day vulnerability/bug in implementation would be a common-mode failure in all blockchains is handled using replaying transactions from \checkpoints, external challengers can provide evidence of malfeasance even when DD/DR fails.
In such a scenario, the transactions since the last \checkpoint (or the disputed point) is replayed in the same order, using software with bug fixes applied, to compute the correct state.

The Oasis network is a blockchain system that utilizes the bare-metal fraud proof approach.
Multiple \commitchain VMs or ``ParaTimes'' are supported.
Transaction ordering is handled via a mempool, and a rotating leader in the compute committee chooses the transaction order in its transaction batch.
This reduces the number of consensus transactions since transaction order finality is not critical when the \commitchain execution is fast, though decoupling can be introduced in later iterations as needed.
The Oasis ParaTime architecture, where the compute layer results are subject to DD/DR before being committed to the consensus layer, is a minimal \commitchain design: the \bridge that validates the \commitchain VMs is baked-in and only supports DD/DR validation, keeping the consensus layer simple.

\dawn{it may be good to include a table to list the different rollup solutions, listing their basechain, validity vs. fraud proof, etc..}

\section{Ideal \Commitchain Design}
\label{sec:IdealL2Design}

The ideal \commitchain design allows efficient (off-chain) verifiable computation with fast finality.
Here, efficiency means:
\begin{itemize}
  \item Reduced replicated computation.
    We do not want to use resources, e.g., use an enormous amount of electricity, when it is unnecessary for the required level of security.
    Reduced replication should also reduce the cost per transaction, making the system more scalable from an economic perspective.
  \item High confidence in results.
    This is the converse to replication: users should be able to trust the results from the verified computation, even though the replication factor might be lower.
  \item High throughput.
    The overall system throughput should be high, so that the system will perform well with a high workload.
  \item Low latency.
    Proposed transactions should not have to wait too long in queues.  Users should have reasonable confidence that once a transaction is accepted, it will have the desired/expected effect.
\end{itemize}

What an ``ideal'' design might look like depends on other factors, such as the expected workload.
For example, as systems become more capable, will applications need to run larger transactions that have a high gas limit?
What is the distribution of execution times?
With a design that separates transaction order finality and state finality, as long as the average transaction processing rate is sufficient, the system should be able to tolerate occasional high execution time transactions, since users will primarily care about transaction order finality, assuming that the expensive transaction do not interfere or interact with most users' contracts' state.

This is still an area of research and some design constraints are unknown.
We present a sketch of an ``ideal'' design below.

\subsection{VM Job Queue and Transaction Order Finality}
\label{sec:VMJobQueue}

The ideal \commitchain design uses ``instant'' finality of a PoS blockchain (e.g. Tendermint, Eth2) to obtain transaction order finality.
That is, transaction proposals are essentially jobs entered into the virtual machine job queue; once entered into the queue, its order is final.

This implies that the transaction callData cannot be validated with full virtual machine semantics, since that requires close coupling between the log and the virtual machine.
One extreme design choice is to allow arbitrary messages to be logged to target the VM's job queue, and rely on the cost of logging to deter denial-of-service attacks.
This means, however, that the cost of VM execution to verify and discard ill-formed message (e.g., unsigned, unparseable, non-existent contract entry points, etc) must be included as part of the cost of message handling.

Message posting has to cost something in order to prevent denial-of-service attacks.
This means some checks must already be done at the \layerone: the message itself must be signed, and it must authorize payment from the EOA to minimally pay for the message posting itself, as well as a (limited) payment authorization for the VM execution (gas limit).

Since signature check is effectively a sunk cost, an alternative is to perform some very simple verification of proposed transactions:
\begin{itemize}
  \item Transaction authorization: valid signature on the proposal.
  \item EOA account has enough tokens to pay for message posting.
  \item (Optional) EOA account has enough tokens to pay for the maximum gas payment, using an lower-bound estimate from the queue of transactions for which state finality has not been reached.
\end{itemize}
The contract call parameters (callData) is just opaque bytes at this point.

All signed transaction proposals with these basic checks are considered ``well-formed transactions'' (WFTs).
Note that we do not perform contract entry point-specific type checks or other per-contract pre-condition checks at the bridge.
Well-formedness does not imply that the message makes sense.
WFTs that do not pass message format checks will simply be aborted with the sender charged a small gas fee: in our formalism, this means that the interpretation is that $t_c$ returns with its input state.
The same thing applies for other pre-conditions: those would be checked by the contract code (e.g., the equivalent of \texttt{require} checks in Solidity) and similarly cause the transaction to be aborted.
This conforms with the decoupled nature of the system design: the \bridge does not know what (possibly new) contracts have been instantiated on the \commitchain, and doing typechecking would be impractical.

Contract invocations in Ethereum-like blockchains are internally structured as essentially remote procedure calls, with a message that must be (partially) parsed to identify the intended entry point and discover the type signature, and then to further parse and typecheck the rest of the message.
While it might be reasonable to say that the message encoding is architecturally fixed and performed by the \commitchain VM, it is feasible---as is done by Solidity for the EVM---to perform this in the contract code itself (in Solidity's case, in the language runtime which performs the RPC dispatch).
Since different contract language runtimes could, in principle, require completely different data serialization formats, performing message format verification in the bridge is infeasible without knowing details about how each contract's RPC message receiver/demultiplexor is implemented.

The simplest design is to view the \bridge as accepting WFTs and require runtime checking for \commitchain-specific, contract-language specific, or contract-specific pre-conditions.

The arrival order at the job queue maintained by the \bridge does not necessarily specify the execution order.
Clearly, as long as a \commitchain allows transaction proposers to offer variable gas fees (as opposed to a fixed market rate as a gating threshold), jobs will need to be sorted by gas price.
And while using gas price as the primary key and arrival order as the secondary key might approximate a ``fair'' order, it allows a frontrunner to trivially creating pseudo-conjugate transactions by proposing a pair of transactions at a slightly higher and slightly lower gas fee.
More research is needed here, especially since implementing an approximate fair scheduler will introduce additional latency to the transaction processing.

\subsection{Data Availability and Garbage Collection}

The availability of callData and state representation are both important.
From the transactions and the ordering, we can reconstruct the state value.
From a finalized state, earlier callData is not needed except for catastrophic failures; from a checkpoint finalized state, earlier callData is no longer accepted.

While the callData is typically stored on-chain, the state representation is too large and cryptographic commitments are used instead.
That doesn't mean that transaction callData \emph{has} to be on-chain.

One scenario is that an external trusted fair scheduling service could be authorized to determine transaction order, with the job schedule stored in an external CAS store with availability guarantees, so that all that the only on-chain data is a cryptographic hash commitment.
The external highly available storage would have to have service-level agreements to maintain the data to remain accessible until at least the data ages beyond checkpoint finality.

The scheduler could run as a decentralized application on a separate blockchain.
In this case, the scheduling algorithm / code is open for inspection and audit and does not have to be trusted; its execution would derive integrity from the blockchain on which it runs, and data availability derives from the availability of this blockchain as well.

Rather than using a separate blockchain, the scheduler could run on the \commitchain VM, with the traditional mempool used only for submitting callData to the scheduler via job-submission transactions.
In such a design, the callData and transaction order would simply be stored as scheduler state in the existing decentralized data availability layer.
The \commitchain \executors would have access to the storage services that holds this and no external highly available data storage service would be needed.

With this kind of integration, the normal state garbage collection mechanism handles reclaiming pre-checkpoint callData: the scheduler can keep the callData in its persistent storage and only remove these entries after a checkpoint has occurred.

\dawn{it'd be good to talk about a decentralized data availability layer so it doesn't rely on a trusted fair scheduler.}
\bsy{Clarified. }

\subsection{State Finality}

State finality determination is where replicated computation occurs, and where efficiency and computation integrity appear to be diametrically opposing goals.

Using reasonable security parameter estimates (see calculations from Appendix~\ref{sec:DDParameters}), we see that if zkSNARK proof generation costs more than about $25\times$ the cost of normal execution, then it will have essentially no replication efficiency advantage, ignoring the costs of the replicated proof verification.
Even if it could be faster, it increases the time to state finality as compared to DD/DR, where the \commitchain committee members execute in parallel.
If we are willing to move away from EVM compatibility for new smart contracts, depending on the contract language and runtime environment design, it should be feasible to allow smart contract to execute at (near) native code speeds with comparatively less engineering effort than proof generation/verification, making the comparison even more one sided.

More importantly, the DD/DR approach is straightforward to analyze and its implementation is much simpler.
Both are critical traits for practical security:  users need to understand why they should trust a system; more importantly, complexity is antithetical to correct, auditable implementations.

Absent a breakthrough in proof generation performance, a DD/DR implemented between commitchain committee and bridge contract provides a good design choice for security, efficiency, and generality.

Note that just as there is the verifier's dilemma for ZK and optimistic rollups, there is a parallel problem in replicated execution / voting schemes.
Execution piggybacking is a potential problem, where an \executor doesn't bother to compute their own resultant state but just re-uses the results from another (hopefully honest) \executor.
To address this, we could try to do something like commit/reveal for results, or have confidential compute requirement for the \commitchain.
Except for trusted computing style mechanisms, such approaches are insufficient: a group of rational \executors can collude to save costs.
Fortunately, as long as there is at least one altruistic \executor to cross verify, the DD/DR approach works.

\subsection{Checkpoint Finality}

DD/DR handles independent failures, e.g., through bribery of key personnel, bugs in custom deployment infrastructures, etc.
The existence of zero day vulnerabilities / bugs mean that there are common-mode failures.
The ``traditional'' way to address such common-mode failures---especially when their exploit was large scaled---is to hard fork and change a checkpoint's hard-wired mapping from a natural name to a state value/representation.

We don't have a better solution, just (hopefully) better terminology / clearer analysis:  checkpoints must lag the current head of the blockchain significantly to permit detection and recovery from common-mode failures, but not so far as to make the long-term high-availability data retention too costly.
This is a balancing act, and the appropriate value depends on the estimated likelihood of common-mode failures, business requirements of dependent contracts, etc.

When to perform a checkpoint should be formalized and made part of the system design and be subject to change by governance.
Storage reclamation / garbage collection and catastrophic fault handling are intimately connected.
The checkpoint state representation and all newer transaction callData must be available to allow recomputation of transaction results in case of common-mode catastrophic faults, but storage for older state representations or callData can be safely reclaimed.
In order for this to work, the servers in the decentralized data availability layer must be aware of \commitchain state, e.g., act as a light client and observe checkpoint determination events themselves or rely on witnesses that relay the information.


\section{Conclusion}
\label{sec:Conclusion}

The main contribution of this paper is that transaction order finality should be viewed as the key for determining system state.
Computing an efficient representation is just an optimization, since state can be reconstituted from a \checkpoint state and those transactions that follows it.
We identify the following shades of finality:
 \textsf{log finality},
 \textsf{transaction order finality},
 \textsf{state finality}, and
 \textsf{\checkpoint finality}.
These notions are useful for reasoning about blockchain design and the design space for error/fault handling, from independent faults due to Byzantine actors to common-mode faults due to zero-day software defects.

Based on considering how and when these finality properties should be achieved, we developed a preliminary design sketch for an ``ideal'' layer 2 system, and discussed some of the trade-offs.

\dawn{what's the difference between log finality and transaction order finality? Clarified.}




\renewcommand{\baselinestretch}{1.0}



{\footnotesize \bibliographystyle{acm}
\bibliography{paper}}

\begin{thebibliography}{10}

\bibitem{albassam2019lazyledger}
{\sc Al-Bassam, M.}
\newblock {LazyLedger}: A distributed data availability ledger with client-side
  smart contracts, 2019.

\bibitem{azouvi2020winkle}
{\sc Azouvi, S., Danezis, G., and Nikolaenko, V.}
\newblock Winkle: Foiling long-range attacks in proof-of-stake systems.
\newblock In {\em Proceedings of the 2nd ACM Conference on Advances in
  Financial Technologies\/} (2020), pp.~189--201.

\bibitem{benet2014ipfs}
{\sc Benet, J.}
\newblock {IPFS}-content addressed, versioned, {P2P} file system.
\newblock {\em arXiv preprint arXiv:1407.3561\/} (2014).

\bibitem{9152675}
{\sc Daian, P., Goldfeder, S., Kell, T., Li, Y., Zhao, X., Bentov, I.,
  Breidenbach, L., and Juels, A.}
\newblock Flash boys 2.0: Frontrunning in decentralized exchanges, miner
  extractable value, and consensus instability.
\newblock In {\em 2020 IEEE Symposium on Security and Privacy (SP)\/} (2020),
  pp.~910--927.

\bibitem{verifiersdilemma}
{\sc Felton, E.}
\newblock The cheater checking problem: Why the verifier’s dilemma is harder
  than you think.
\newblock Accessed 21/08/2021,
  \url{https://medium.com/offchainlabs/the-cheater-checking-problem-why-the-verifiers-dilemma-is-harder-than-you-think-9c7156505ca1}.

\bibitem{190890}
{\sc Heilman, E., Kendler, A., Zohar, A., and Goldberg, S.}
\newblock Eclipse attacks on bitcoin{\textquoteright}s peer-to-peer network.
\newblock In {\em 24th {USENIX} Security Symposium ({USENIX} Security 15)\/}
  (Washington, D.C., Aug. 2015), {USENIX} Association, pp.~129--144.

\bibitem{hopwood2016zcash}
{\sc Hopwood, D., Bowe, S., Hornby, T., and Wilcox, N.}
\newblock Zcash protocol specification.
\newblock {\em GitHub: San Francisco, CA, USA\/} (2016).

\bibitem{kalodner2018arbitrum}
{\sc Kalodner, H., Goldfeder, S., Chen, X., Weinberg, S.~M., and Felten, E.~W.}
\newblock Arbitrum: Scalable, private smart contracts.
\newblock In {\em 27th $\{$USENIX$\}$ Security Symposium ($\{$USENIX$\}$
  Security 18)\/} (2018), pp.~1353--1370.

\bibitem{kelkar2020order}
{\sc Kelkar, M., Zhang, F., Goldfeder, S., and Juels, A.}
\newblock Order-fairness for byzantine consensus.
\newblock In {\em Annual International Cryptology Conference\/} (2020),
  Springer, pp.~451--480.

\bibitem{luu2015demystifying}
{\sc Luu, L., Teutsch, J., Kulkarni, R., and Saxena, P.}
\newblock Demystifying incentives in the consensus computer.
\newblock In {\em Proceedings of the 22nd ACM SIGSAC Conference on Computer and
  Communications Security\/} (2015), pp.~706--719.

\bibitem{zkEVM}
{\sc {Matter Labs}}.
\newblock {zkSync} 2.0: Hello ethereum!
\newblock
  \url{https://medium.com/matter-labs/zksync-2-0-hello-ethereum-ca48588de179}.

\bibitem{InvestopediaFrontRunning}
{\sc Mitchell, C.}
\newblock Front-running.
\newblock \url{https://www.investopedia.com/terms/f/frontrunning.asp}.
\newblock Accessed: 2021-09-01.

\bibitem{MEV}
{\sc Noyes, C.}
\newblock {MEV} and me.
\newblock Accessed 08/09/2018, \url{https://research.paradigm.xyz/MEV}.

\bibitem{teutsch2019scalable}
{\sc Teutsch, J., and Reitwie{\ss}ner, C.}
\newblock A scalable verification solution for blockchains.
\newblock {\em arXiv preprint arXiv:1908.04756\/} (2019).

\end{thebibliography}

\appendix
\section{Discrepancy Detection Security Parameters}
\label{sec:DDParameters}

The key security parameter for the discrepancy fast-path approach is the size of the primary committee used for the DD protocol.
The size of the slow path DR backup committee depends on the DR protocol chosen, but it can be chosen generously, since the DR protocol should be used extremely infrequently.
Any existing state verification protocol can be used, e.g., consensus based on majority voting, 2/3 super-majority, etc.

For the DD primary committee, the security assumptions are that faults are independent, e.g., due to installation-specific configuration errors, personnel security, bribery, etc; and that at most some fraction of the available participants are faulty.
We will use a random selection to choose a subset of the available participants to serve in the primary committee.
Because we are looking for any discrepancy in the reported output state, this is \emph{not} a stake-weighted voting scheme.

Let $T$ denote the total number of available participants, and $B$ denote the number of faulty or Byzantine participants.\footnote{If we wish to use the BAR model, instead of $B$ we would use $B+R$ as the worst case to assume that all rational participants could be bribed etc, by the virtue of arbitrary token creation/transfers.}
We need to pick $C$, the size of the primary committee.

Our goal is to make the probability that the entire primary committee are faulty is negligibly small.
An acceptable level would be small enough such that the expected number of committee formations needed before an all-faulty committee is chosen is enormous, so that even with a high worst-case rate of committee selections, many lifetimes must pass before such an event would take place.

The combinatorics is reasonably straightforward.
The number of all-faulty committees is $\binom{B}{C}$, and the number of possible committees is $\binom{T}{C}$, so the expected number of committee selections before an all-faulty committee is selected is
\[W = \frac{\binom{T}{C}}{\binom{B}{C}}\]

Since DR schemes cannot work unless there is at least an honest majority, we will use $B\approx T/2$ as worst-case parameters, e.g., $T=100$ and $B=49$.
With a committee size of $C=25$, this is an expected $W=3.837\cdot 10^{9}$ committee selections.
Assuming a committee selection rate of $1,000$ per hour, this works out to about $437$ years before an all-faulty committee is encountered.
This ramps up quickly: at $C=26$, it works out to be about $1,368$ years; at $C=30$, it is about $177,740$ years.
Obviously if we were to assume $B \approx T/3$, the primary committee can be even smaller for the same level of security.

An interesting observation is that DD could work with a pool of participants that are less trustworthy than would be feasible with other protocols, i.e., where honest nodes are \emph{not} a majority, as long as the DR participant pool is acceptable (e.g. a majority are honest for when DR uses honest majority).
Estimating whether participants are corruptible is difficult, obviously, and it is not clear why we might have to work with a faulty-majority participant pool, though the possibility is intriguing.
A possibility is to require periodic external security audits to participate in the DR pool, and have less stringent requirements to participate in the DD pool.

\end{document}